\documentclass[final,2p,times,12pt]{elsarticle}

\usepackage{graphics}  
\usepackage{lscape,tipa}
\usepackage{multirow,array,bm,enumerate}
\usepackage{amssymb,epsfig,amsmath,comment,hyperref,verbatim}
\usepackage{subfigure,setspace}
\usepackage{textcomp}
\usepackage{url}
\usepackage{color}

\DeclareMathOperator*{\argmax}{arg\,max}

\usepackage{amsmath}
\usepackage{comment}
\usepackage{tikz}
\usepackage{tikz}
\usetikzlibrary{arrows,patterns}
\usetikzlibrary{positioning}

\usepackage{lineno,hyperref}

\journal{}

\begin{document}

\begin{frontmatter}

\title{Direction of arrival estimation from distant microphone data using single frequency filtering}

\author{Sushmita Thakallapalli$^1$, Sudarsana Reddy Kadiri$^2$, Nilesh Madhu$^3$, and Suryakanth V Gangashetty$^4$}
\address{$^1$Speech Processing Laboratory, International Institute of Information Technology, Hyderabad, India\\$^2$Signal Analysis and Interpretation Laboratory, University of Southern California, Los Angeles, USA\\$^3$IDLab, Dept. Electronics \& Information Systems, 
Ghent University - imec, Belgium\\ $^4$Koneru Lakshmaiah Education Foundation, Vaddeswaram, Guntur District, Andhra Pradesh, India}

\cortext[mycorrespondingauthor]{Corresponding author (Sudarsana Reddy Kadiri; sudarsana.kadiri@aalto.fi)}

\begin{abstract}
In distant microphones, broadband (BB) methods for direction-of-arrival (DoA) estimation are more suitable than narrowband (NB) methods. Due to the aggregation of their optimization function across all frequency bands, BB estimators are robust to spatial aliasing, a known problem in processing distant microphone data. In NB methods, DoA estimation is performed by utilizing \textit{local} information in each frequency band and hence the estimation is affected by spatial aliasing. However, unlike BB methods, NB methods exploit frequency sparsity to estimate the DoAs of \textit{multiple speakers} in a \textit{single time frame}. In this article, a method to improve the robustness of a NB DoA estimator to spatial aliasing is developed. The proposed method is based on cross-correlation of speech-present time-frequency regions obtained by single frequency filtering (SFF) of the microphone signals. The SFF spectrum is chosen because SFF components have regions of high signal-to-noise ratio both in time and frequency and because speech and non-speech discrimination is robust to degradations in the SFF domain. The proposed NB estimator is compared to four state-of-the-art estimators (one NB and three BB) using detection and accuracy metrics on simulated and real-world data in different reverberation and noise conditions. The results show that in all the environments, the SFF-based NB approach outperforms the state-of-the-art NB approach. Furthermore, the performance of the SFF-based approach is better than some of the BB estimators.
\end{abstract}


\end{frontmatter}

\section{Introduction}
Networks with large microphone spacings (0.15 - 6 m apart) have gained attention in applications such as source separation, signal enhancement and tracking  \cite{2016_adhoc_loc, 2016_adhoc_loc1}. In these applications, the first and primary step is sound source localization (SSL). In most SSL approaches, direction-of-arrival (DoA) is estimated by exploiting some form of cross-correlation of signals across microphones.

\par
The DoA estimation can be performed using broadband (BB) methods \cite{1976_GCC_knapp, 2001_dibiase_WSRP, 2000_nbloc_liu} or narrowband (NB) methods \cite{2004_nbsrp_hist1, 2006_mask_Mouba, 2003_nb_hist3_on_spatial_cues, 2008_nbsrp_hist2,madhu08mog}. While NB methods exploit sparsity and disjointness of speech in a time-frequency (T-F) representation, BB approaches exploit time sparsity only  \cite{2021_srp_nmf}. NB methods first compute the DoA of the mixture signal locally in each T-F bin after which they localize the peaks of the resulting histogram. BB methods, on the other hand, compute the DoA by averaging the locally computed functions of the DoA across frequencies. Due to this averaging, BB methods are robust to spatial aliasing. However, since BB methods do not exploit frequency domain sparsity, their use leads to poor DoA estimates in the case when the speech signal waveform has no prominent temporal components (such as in soft speech). 
On the other hand, NB methods can estimate DoAs of multiple speakers in a single frame but these methods are affected by spatial aliasing.
To improve DoA estimation accuracy of NB techniques, this study proposes a method, which computes the DoA locally using those T-F regions where speech is present. The DoA estimation in speech-present segments leads to accurate DoA estimates, while the DoA estimation in segments, where speech is absent or where signal-to-noise ratio (SNR) is poor, results in inaccurate DoA estimates. 

\par
 In this paper, we propose a DoA estimation method using voiced speech recorded by a distant microphone. To compute DoA of a sound source in signal regions where (voiced) speech is present, both a robust voice activity detection (VAD) algorithm 
 as well as a T-F representation are needed. We develop 
 a VAD algorithm and a T-F representation based on single frequency filtering (SFF) \cite{sff_2015_aneeja}. Selecting SFF as the basis of VAD and the T-F representation is justified, respectively, by the following rationales. First, a SFF-based VAD method was reported in \cite{sff_2015_aneeja} to be robust with respect to various additive noise degradations. 
The key idea proposed in  \cite{sff_2015_aneeja} is to use variance of spectral information across frequencies to discriminate speech from non-speech. The variance is higher for speech compared to different types of noise signals. In this paper, spectral flatness, which is a measure of variance, is used for VAD. The idea of using spectral flatness for VAD has also been studied in \cite{drygajlo_2001_spectral_flatness, 2009_Nilesh_spectralFlatness}. Second, spectral envelopes of the SFF-based TF representation have regions of high SNR both in the time and frequency domain \cite{sff_2015_aneeja}. 
In voiced speech, high SNR regions are located around glottal closure instants (GCIs), where the acoustical excitation of the glottal source is most prominent. 
By identifying speech samples in high SNR regions in voiced segments, cross-correlations are computed. This results in robust DoA estimates because speech samples in the vicinity of GCIs are of high SNR and therefore less affected by degradations \cite{2005_Hilbert_yegna}. Furthermore, SFF offers a good trade-off between time and frequency resolution. The main highlights of this study are as follows:

\begin{itemize}
    \item We propose a new SFF-based NB DoA estimator, which takes advantage of signal regions of high SNR during voiced speech segments.
    \item We conduct a systematic comparison of the SFF-based NB DoA estimator with a well-known NB DoA estimator and widely used three BB DoA estimators. 
    \item We report experiments based on both simulated as well as real-world data in noisy and reverberant conditions.
\end{itemize}

The paper is organized as follows: Prior work in the topic is described in Section \ref{sec:prevWork}. In Section \ref{sec:proposedMethod}, the SFF-based DoA estimation approach is presented. The experimental setup is described in Section \ref{sec:exptSetup} and the results are reported and discussed 
in Section \ref{sec:results}. Finally, the summary and conclusions of the study are given in Section \ref{sec:conclusion}. 

\section{Prior work}
\label{sec:prevWork}
SFF-based TF representations have been successfully applied in different speech processing applications such as detection of speech regions \cite{sff_2015_aneeja}, GCI/epoch extraction \cite{2017SFFEpoch,kadiri2020determination,AneejaKY18}, dysarthic  speech detection \cite{2019_icassp_krishna}, and time-delay estimation \cite{2019_sff_loc_sudarsana}. Furthermore, high SNR regions of 
mixtures (signals captured by multiple microphones) were used for time delay estimation in \cite{2005_Hilbert_yegna} and SNR weighting of localization approaches was studied in \cite{2019MaskGccPhat} and in \cite{2016PreFilteringGCCPhat}. 
The present study extends these ideas to a SFF-based TF representation for (i) VAD, and (ii) DoA estimation in voiced regions of speech signals. 
\\ \indent SFF was studied in the sound source DoA estimation in \cite{2019_sff_loc_sudarsana}, where 
cross-correlation was computed at all time instants between the SFF spectral envelopes from different microphones to obtain DoA estimates. This approach has the following two drawbacks. First, cross-correlation in signal regions where speech is absent leads to false DoA estimates. Second, the computational complexity of the method is high because correlation is performed at all frequencies and at all time instants. To tackle these issues, the focus of this study is on performing cross-correlation only in specific regions of voiced speech in order to get DoA estimates that are accurate and robust to degradations. 
Furthermore, we evaluate the proposed approach in the current study using a more versatile setting compared to \cite{2019_sff_loc_sudarsana}: the simulated dataset of the current study evaluates the DoA estimator in conditions of severe reverberations and very low SNR whereas the experiments in \cite{2019_sff_loc_sudarsana} were based on 
recordings in laboratory conditions, which lack the diversity of room acoustic conditions. In addition, since the proposed SFF-based DoA estimation approach is a NB method, it is compared in this study with a state-of-the-art NB approach,
Steered Response Power with Phase Transform Weighting (SRP-PHAT) as described in \cite{srp_nmf}, along with various WB methods including \text{Hilbert} Envelope of Linear Prediction residual (HE-LP) \cite{2005_Hilbert_yegna}, Generalized Cross Correlation (GCC) \cite{1976_GCC_knapp} and Generalized Cross Correlation with Phase Transform Weighting (GCC-PHAT) \cite{1976_GCC_knapp}. In \cite{2019_sff_loc_sudarsana}, such multiple comparisons were not studied. 

\section{Using SFF in Sound Source D\lowercase{o}A estimation}
\label{sec:proposedMethod}
In this section, the benefits of the SFF representation over the conventional short-time Fourier transform (STFT) representation are discussed and the SFF-based source DoA estimation algorithm is described. 
\subsection{ Time-frequency representation by narrowband filtering}
\label{subsec:whySFF}
To compute the SFF-based TF representation, the speech signal is filtered with a narrowband filter which has a single complex-valued pole. 
The transfer function of this filter is given by:
\begin{equation}
    H(z)= \frac{z}{z+r},
    \label{eq:transfucn}
\end{equation}
 where $r= |r|e^{j\omega_r}$ is the pole location. For a stable filter, $|r|<1$. The value of $|r|$ determines the filter bandwidth ($B$): 
 the larger the value of $|r|$, the narrower the bandwidth. Since a narrow bandwidth gives good frequency resolution, $|r|$ is chosen close to 1.  The chosen value of $r$ in this study is 0.995. 
 The center frequency of the filter is set to half of the sampling frequency ($f_s$) corresponding to $\omega_r=\pi$.
 
 \par
\begin{figure*}[ht!]
  \centering
  \includegraphics[width=14cm,height=3.8cm]{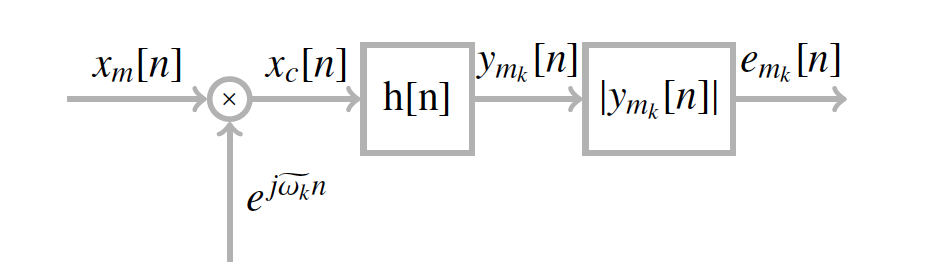}
  \caption{Block diagram of the computation of the spectral envelope $e_{m_k}[n]$ for the $k$th 
  frequency from the $m$th 
  microphone signal.}
  \label{fig:sffAsFilter}
\end{figure*}

The SFF spectral envelope can be obtained at any desired frequency $f_k$. Let $[f_1, f_2, ..., f_K]$ be the frequency components of interest. To estimate the SFF spectral envelope at frequency $f_k$, we first assign $\tilde{f_k} = \frac{f_s}{2} - f_k$. Subsequently, the spectral envelope $e_{m_k}[n]$ at $f_k$ in the $m$th microphone channel is estimated as shown in Figure \ref{fig:sffAsFilter}, where $x_m[n]$ is the input signal at the $m$th microphone, $\widetilde{\omega_k} = 2\pi \widetilde{f}_k$, $x_c[n]$ is the complex output of the multiplier, $h[n]$ is the impulse response of the filter defined in 
Equation ~\eqref{eq:transfucn} and $y_{m_k}[n]$ is the complex output of the filter.

\begin{figure*}[t!]
  \centering
  \includegraphics[width=\textwidth,height=7cm]{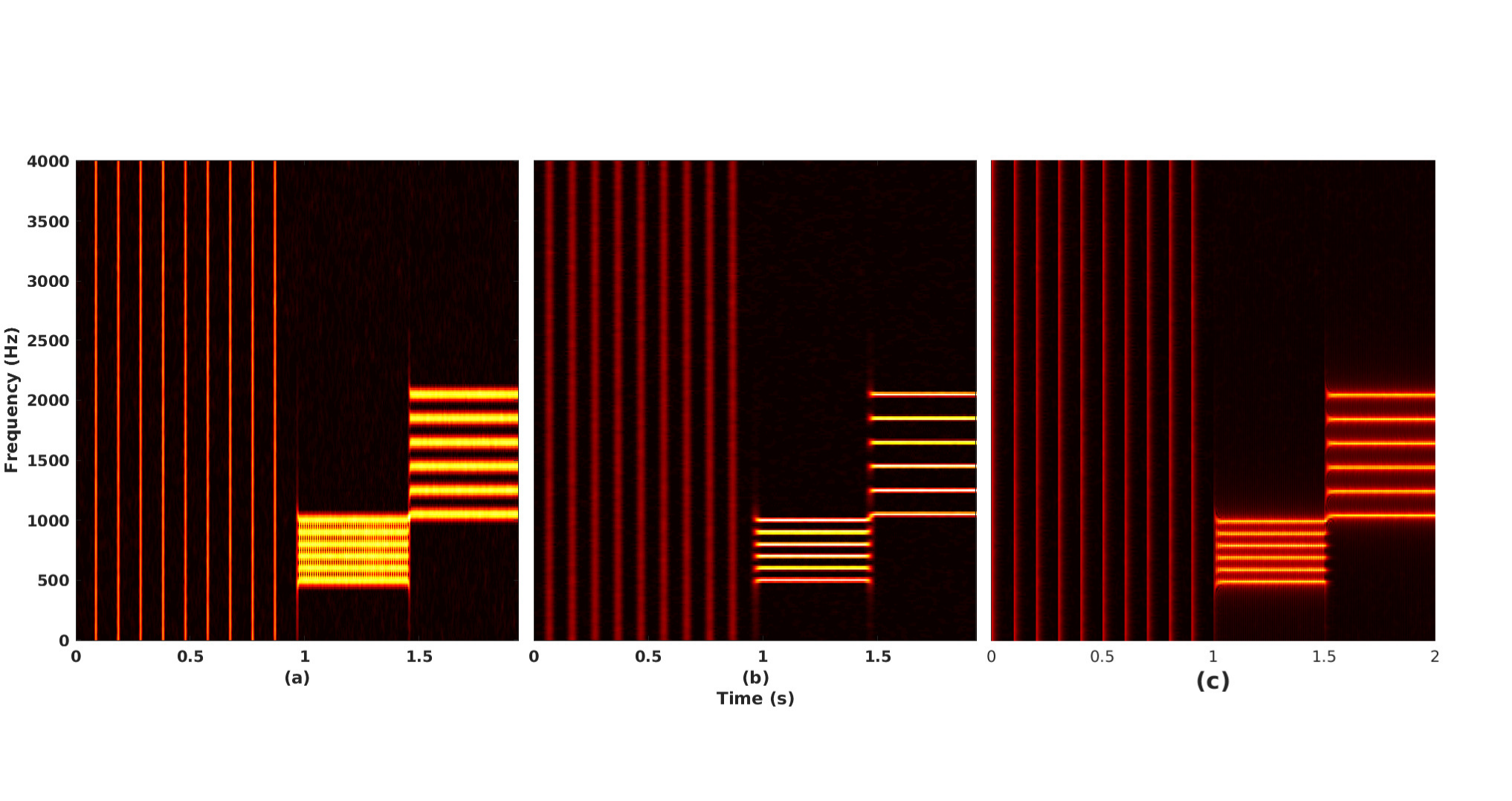}
    \vspace{-1.2cm}
  \caption{TF representations for a synthetic signal consisting of impulses and narrowband components. Panels (a) and (b) were computed using STFT with a frame size of $20$ $ms$ and $64$ $ms$, respectively, using a hop size of one sample. Panel (c) shows the SFF-based TF representation computed using $r = 0.992$. In all panels, the number of frequency components between $0$ and $\frac{f_s}{2}$ is 256 (adapted from \cite{SpectralFeat_NCC}).}
  \label{fig:STFT_SFF}
\end{figure*}


\indent In Figure \ref{fig:STFT_SFF}, analysis of a synthetic signal consisting of a sequence of impulses and narrowband signals is depicted based on the experiments reported in \cite{SpectralFeat_NCC}. Panels (a) and (b) depict the results obtained using STFT computed with the 512-point FFT using a short (20 ms) and long (64 ms) Hamming window, respectively, and a hop size of one sample. Panel (c) shows SFF of the same signals with $r = 0.992$ and $K = 256$. It can be observed that in panel (a) the temporal resolution is good while spectral leakage between the bands is clearly visible. On the other hand, in panel (b), closely spaced spectral components are visible but at the cost of decreased time resolution. In panel (c), good spectral and temporal resolution (at impulses) can be observed. SFF thus gives a good trade-off between frequency and time resolution for $0.95 \leq |r| \leq 0.995$ \cite{2019_icassp_krishna,ChennupatiKY18}.     
\subsection{The  SFF-based NB sound source D\lowercase{o}A estimator} 
\label{subsec:algoDetails}
In this section, the SFF-based approach for the DoA estimation of sound sources from degraded stereo mixtures is described. The basic idea is to compute the DoA estimates by cross-correlating the SFF spectral envelopes at all frequencies using signal regions where speech is present. 
These regions of high SNR contain information of the excitation impulse sequence which are robust to degradations \cite{2019_sff_loc_sudarsana, 2005_Hilbert_yegna}. Furthermore, these regions are obtained with good resolution as demonstrated in subsection \ref{subsec:whySFF}. Therefore, the DoA estimates obtained from these high SNR regions are more likely to be accurate. Since these high SNR regions are primarily present in voiced speech segments, a robust SFF-based VAD is used. \\ \indent To detect voiced regions in SFF envelopes, a spectral flatness measure is computed. Spectral flatness is higher in regions where speech is absent compared to regions where speech is present. An illustration of the use of the spectral flatness measure for VAD is shown in Figure \ref{fig:sfmeasure}. In Figure \ref{fig:sfmeasure}, panel (a) shows a noisy speech signal corrupted with white noise at SNR of $-5$ dB and panel (b) depicts the corresponding clean speech signal and spectral flatness estimated from the signal shown in panel (a). As observed from panel (b), the low values (troughs) in $\delta[n]$ coincide with speech segments.  While spectral flatness of speech is small due to its correlated nature, this measure is higher for uncorrelated noise \cite{2009_Nilesh_spectralFlatness}. In distant microphones, noise can be assumed to be uncorrelated 
due to reduced spatial coherence between the microphones \cite{2007_Habets_diffuse_noise}.  

The steps involved in the proposed SFF-based sound source DoA estimation approach are as follows:
\begin{enumerate} 
    \item SFF spectral envelopes of mixture signals $x_m[n]$ are estimated using the SFF technique as described in subsection \ref{subsec:whySFF}. 
    \item Voiced speech segments are identified using spectral flatness $\delta[n]$ across the SFF frequency components in 
    all microphone channels. 
    \begin{itemize}
        \item $\delta[n]$ is calculated by dividing the geometric and arithmetic means of the normalized and squared spectral envelopes according to Equation ~\eqref{eq:sf}.  
   
    \begin{equation}
    \delta[n] = \frac{\sqrt{\prod_{k=1}^{K}e_{m_k}^2[n]}}{\frac{1}{K} \sum_{k=1}^{K}e_{m_k}^2[n]},
    \label{eq:sf}
\end{equation}
where $m$ refers to the microphone, (i.e., $m = 1$ or $2$), $k$ refers to the $k$th frequency channel ($f_k$). $e_{m_k}[n]$ denotes the spectral envelope at frequency $f_k$ of the $m$th microphone. 
    \end{itemize}

\begin{figure*}[htb]
\begin{minipage}[b]{1.0\linewidth}
  \centering
  \centerline{\includegraphics[scale=0.95]{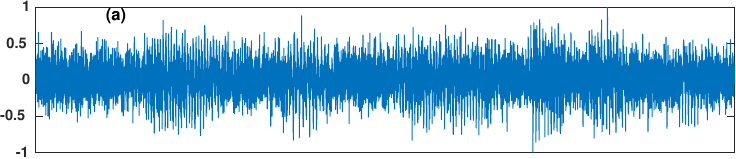}}
\end{minipage}
  \vspace{-0.1cm}
\begin{minipage}[b]{1.0\linewidth}
  \centering
  \centerline{\includegraphics[scale=0.95]{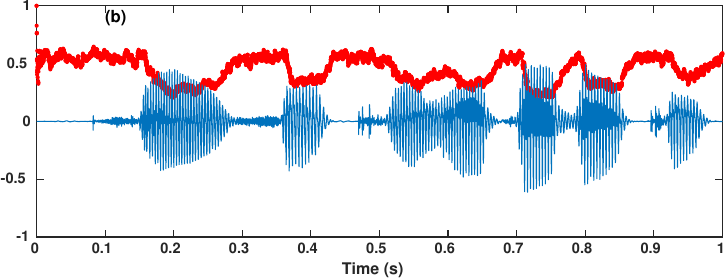}}
\end{minipage}
\caption{An example of computing spectral flatness for a sentence. Panel (a) shows the speech signal corrupted with additive white noise at SNR of -5 dB. Panel (b) shows the spectral flatness (red curve) estimated using the SFF spectrum from the noisy signal shown in panel (a). The original clean signal (blue curve) is shown as reference in  panel (b). 
}
\label{fig:sfmeasure}
\end{figure*}
\item A few lowest valleys in the spectral flatness are chosen by a peak picking algorithm. Subsequently, a few hundred samples around the valley locations are extracted from the envelopes at all frequencies. 
\begin{itemize}
    \item Let $t_v$ be the time index of a valley location in $\delta[n]$. In the SFF approach, 30 such locations are chosen with a minimum separation of 500 samples between the adjacent valleys. The choice of these numbers is not critical. 
    Considering 30 valleys enables getting multiple evidence for the DoA estimation. A distance of 500 samples (50 ms) between the adjacent valleys ensures that the valleys are picked across the signal and not from a restricted region.
    \item Let $y_{m_{k}t_{v}}[n]$ as in Equation ~\eqref{eq:seg} represent a short segment of the $k$th spectral envelope in the $m$th microphone channel containing samples around $t_v$.
\begin{equation}
    y_{m_kt_v}[n] = [e_{m_k}[t_v-N], . . ., e_{m_k}[t_v], . . . e_{m_k}[t_v+N]], 
    \label{eq:seg}
\end{equation}
where $N=250$. Thus $y_{m_kt_v}[n]$ has 501 samples (50 ms).
\end{itemize}
\item The segments $y_{1_kt_v}[n]$ and $y_{2_kt_v}[n]$ are cross-correlated according to Equation ~\eqref{eqn:Eqn7}. 
\begin{equation}
	\label{eqn:Eqn7}
	c[\tau] ={\sum_{n}{y_{1_kt_v}[n]y_{2_kt_v}[n+\tau]}}, 
	-\tau_{max}\leq \tau \leq \tau_{max},
\end{equation}
where $\tau_{max} = 29$, since the maximum delay with a microphone separation of 1 m and with $f_s$ = 10 kHz is 29 samples. Subsequently, the peak location indicates a \textit{local} DoA estimate.
\item 
Assuming that the number of speakers is known, the location of top peaks in the histogram of the \textit{local} DoA estimates (obtained from all valley locations and across all frequencies) are the final location estimates of the sources. 

\end{enumerate}    

\section{Experimental Setup} 
\label{sec:exptSetup}
The performance evaluation of the SFF-based source DoA estimation  approach described in Section \ref{sec:proposedMethod} was performed using simulated data and real-world recordings and compared with four baseline techniques. In this section, the details of the datasets, the baseline DoA estimators used for comparison, the parameters chosen for the DoA estimators and the performance metrics are described.

\subsection{Simulated dataset} 
\label{subsec:simulated}
Two omnidirectional microphones, 1 m apart, were placed in a simulated rectangular room (dimensions: $5.6 \times 4.5 \times 2.6 $ m) as in \cite{2005_Hilbert_yegna}. The details of the simulator are described in \cite{2016roomsim}. Clean speech data of four male and four female speakers from the Signal Separation and Evaluation Campaign (SiSEC) \textit{dev1} database were used as the source speakers \cite{sisec}. A source was placed at a distance of 2 m using seven angles from the centre of the array. The seven DoAs were as follows: $30^o, 50^o, 70^o, 90^o, 110^o, 130^o$ and $150^o$. The room impulse response was calculated for each of these seven positions of the source and the microphone pair and for four different reverberation times (0.0 s, 0.1 s, 0.2 s, 0.3 s). Thus for each source, 28 mixture files were generated, resulting in 224 combinations (i.e. mixture files) of speakers, speaker locations and reverberation times. In addition to the simulated reverberation data, simulated noisy speech were also generated. Noisy speech mixtures were generated by adding white noise from the NOISEX database \cite{noisex92} using four SNRs (-10 dB, -8 dB, -5 dB, 0 dB).

\subsection{Real-world dataset}
\label{subsec:real}
The real-world dataset chosen was 
the development set (both \textit{dev1} and \textit{dev2}) of SiSEC \cite{sisec, sisecDetails}.
The SiSEC live speech recording datasets consist of ten 10 s stereo mixtures of three female and four male speakers. The mixtures were recorded with reverberation times of 130 ms and 250 ms using a microphone separation of 1 m. The ground-truth DoAs of the sources are provided in the dataset \cite{sisec}.

\subsection{\label{subsec:Compare_approaches}Baseline D\lowercase{o}A Estimators} This section describes the baseline DoA estimators that were compared with the proposed SFF-based DoA estimator. The following four methods were selected as the  baseline estimators to be used in the comparison: \text{Hilbert} Envelope of Linear Prediction residual (HE-LP) \cite{2005_Hilbert_yegna}, Generalized Cross Correlation (GCC) \cite{1976_GCC_knapp}, Generalized Cross Correlation with Phase Transform Weighting (GCC-PHAT) \cite{1976_GCC_knapp}, and Narrowband Steered Response Power with PhaseTransform Weighting (NB-SRP-PHAT) \cite{dibiase2000high}. 

\subsubsection{\label{subsub:HE}\text{Hilbert} Envelope of Linear Prediction Residual (HE-LP)} In order to estimate the DoAs, the \text{Hilbert} envelopes of the linear prediction (LP) residual (\cite{LP_1975}) of the microphone signals are cross-correlated \cite{2005_Hilbert_yegna}. 
The \text{Hilbert} envelope of the LP residual signal $\textit{l}[n]$ is defined as:
\begin{equation}
    h[n] = \sqrt{\textit{l}^2[n] + \textit{l}_h^2[n]},
\end{equation}
where $\textit{l}_{h}[n]$ is the \text{Hilbert} transform of $\textit{l}[n]$.
Let $h_1[n]$ and $h_2[n]$ denote the \text{Hilbert} envelopes of the LP residuals of the speech signals at two microphones.  
Let $h_{1b}[n]$ and $h_{2b}[n]$ be two corresponding frames each with a length of $N$ samples. The cross-correlation function of these frames is given by: 

\begin{equation}
	\label{eqn:hilbert}
	\mathcal{J}_{\text{HE-LP}}[\tau(\theta),b] ={\sum\limits_{n=0}^{{N}-1}}{h_{1b}[n]h_{2b}[n+\tau(\theta)]}, 
\end{equation}
where $-T_{max}\leq \tau(\theta) \leq T_{max}$ and $T_{max}$ is the maximum time-delay between the given pair of microphones. The cross-correlation function is estimated at each frame $b$ where $0\leq b \leq B$. Each block begins at $b\Delta T$. The peak location in $\mathcal{J}_{\text{HE-LP}}[\theta,b]$ at each time frame is estimated as:
\begin{equation}
    \hat{\theta}(b) =  {\argmax_{\theta}}~~ \mathcal{J}_{\text{HE-LP}}[\tau(\theta),b].
\end{equation}
The peak locations in the histogram plot of $\hat{\theta}(b)$ for $0\leq b \leq B$ correspond to the source DoA estimates. Since the correlation is performed in the time domain, HE-LP is a BB method.

\subsubsection{\label{subsub:GCC}Generalized Cross Correlation (GCC)}
Generalized Cross Correlation is defined as the cross-correlation of filtered versions of the signals  \cite{1976_GCC_knapp}. The cross-correlation function $\mathcal{J}_{cc}[\tau(\theta),b]$ is obtained by frequency domain operations as follows: 

\begin{equation}
   \label{eq:gcc_freq}
    \mathcal{J}_{GCC}[\tau(\theta),b] =   \sum_{k}{W(k){X_{1}[k,b]X^*_{2}[k,b]}e^{-j2\pi f_k \tau(\theta)}},
\end{equation}
where $W(k)$ is the weight function, $X_{1}[k,b]$ and $X_{2}[k,b]$ are the STFT of $x_1[n]$ and $x_2[n]$, respectively, and  $f_k$ is the frequency (in Hertz) corresponding to the $k$th bin. Since the correlation involves summing across frequencies, GCC is a BB method.  Different weight functions were studied in~\cite{1976_GCC_knapp}. In this study, two variants of GCC are used as baseline DoA estimators. The first one, referred to as GCC, is obtained by $W(k)=1$ and the second one, referred to as GCC-PHAT, is described in Section~\ref{subsub:GCCPHAT}. At each time frame $b$, the peak location in $\mathcal{J}_{\text{GCC}}[\tau(\theta),b]$ is estimated, indicating the DoA of an active speaker in that time-frame as: 
 \begin{equation}
   \label{eq:bb_gcc2}
    \hat{\theta}(b) =  {\argmax_{\theta}}~~ \mathcal{J}_{\text{GCC}}[\tau(\theta),b].
\end{equation}
The peak locations in the histogram plot of $\hat{\theta}(b)$ for $0\leq b \leq B$ correspond to the source DoA estimates.

\subsubsection{\label{subsub:GCCPHAT}Generalized Cross Correlation with Phase Transform Weighting (GCC-PHAT)}
In GCC-PHAT, the cross-correlation function $\mathcal{J}_{\text{GCC-PHAT}}[\tau(\theta),b]$ of the signals at two microphones, $x_1[n]$ and $x_2[n]$, is given by:

\begin{eqnarray}
   \label{eq:bb_gcc_phat}
    \mathcal{J}_{\text{GCC-PHAT}}[\tau(\theta),b]   =\text{GCC-PHAT}(x_1[n],x_2[n]) \nonumber \\  =\sum_{k}{\frac{X_{1}[k,b]X^*_{2}[k,b]}{|X_{1}[k,b]||X^*_{2}[k,b]|}e^{-j2\pi f_k \tau(\theta)}}
\end{eqnarray}
where $X_{1}[k,b]$ and $X_{2}[k,b]$ are the STFT of $x_1[n]$ and $x_2[n]$, respectively, and $f_k$ is the frequency (in Hertz) corresponding to the $k$th bin.
At each time frame $b$, the peak location in $\mathcal{J}_{\text{GCC-PHAT}}[\tau(\theta),b]$ is estimated, indicating the DoA of an active speaker in that time frame as: 
 \begin{equation}
   \label{eq:bb_gcc_phat2}
    \hat{\theta}(b) =  {\argmax_{\theta}}~~ \mathcal{J}_{\text{GCC-PHAT}}[\tau(\theta),b].
\end{equation}
The peak locations in the histogram plot of $\hat{\theta}(b)$ for $0\leq b \leq B$ correspond to the source DoA estimates.

\subsubsection{\label{subsub:SRPPhat}Narrowband Steered Response Power with Phase Transform Weighting (NB-SRP-PHAT)}
In NB-SRP-PHAT, the target DoA is found from the peaks of  the histogram of the local DoA estimates at each TF bin. The DoA estimate is obtained at a time frame $b$ and frequency bin $k$ as follows:
\begin{equation}
   \label{eq:NB-SRP-PHAT}
    \mathcal{J}_{\text{NB-SRP-PHAT}}[\tau(\theta),k,b] =   \frac{X_{1}[k,b]X^*_{2}[k,b]}{|X_{1}[k,b]||X^*_{2}[k,b]|}e^{-j2\pi f_k\tau(\theta)}.
\end{equation}
At each TF bin, the peak location in $\mathcal{J}_{\text{NB-SRP-PHAT}}[\tau(\theta),k,b]$ is estimated, indicating the DoA of an active speaker in that TF bin as: 
 \begin{equation}
   \label{eq:NB-SRP-PHAT2}
    \hat{\theta}(k,b) =  {\argmax_{\theta}}~~ \mathcal{J}_{\text{NB-SRP-PHAT}}[\tau(\theta),k,b].
\end{equation}
The DoA estimates at each TF bin ($\hat{\theta}(k,b)$) are pooled by a histogram. The peak locations in the histogram are considered as source DoA estimates.

\subsection{Parameters used in D\lowercase{o}A estimators}
\label{subsec:baselines}
The use of the baseline DoA estimators described in the previous sub-sections calls for setting parameters for proper comparison with the proposed SFF-based DoA estimator. 
For all DoA estimators, $f_s$ = 10 kHz is considered and the DoA estimation is performed using 2.5 s of data.  The maximum time difference of arrival (TDoA) for microphones that are 1 m apart from each other is 2.9 ms (29 samples). The number of TDoAs/DoAs used in all the approaches is 59, corresponding to a resolution of 3 degrees. \\ \indent To have a fair comparison, the temporal resolution is equal in all DoA estimators and only voiced speech segments of high SNR are used in the DoA estimation. \begin{itemize}
    \item In SRP-PHAT and GCC-PHAT, complex spectrograms are estimated with 512-point FFT using the Hann window with a frame duration of 50 ms and a hop size of 1 sample.
    \item In GCC, GCC-PHAT and SRP-PHAT, the instants of high SNR are detected just as in the SFF approach, but using the STFT spectrum. In HE-LP, the corresponding instants detected by the SFF approach are used. 
    \end{itemize} 
    After the detection of the instants of high SNR, 300 (30 ms) samples to the left and right of the 30 deepest valleys (corresponding to the minimum distance between valleys equalling 500 samples) are used for correlation. Correlation is computed using  
    50 ms time-frames.

\subsection{Performance Metrics}
\label{subsec:Performance Metrics} 
Two performance metrics, root mean square error (RMSE) in degrees and the percent of missed detections (MD), are considered. If the estimated DoA is not within $\pm{5}^o$ of the ground truth angle, it will be counted as a missed detection. The better the system performance, the lower the value of both RMSE and MD.

\section{ Results and Discussion}
\label{sec:results}


This section describes the results of the comparison between the various DoA estimators. The results are first reported for the simulated dataset (Tables \ref{tab:rev} and \ref{tab:noise}) and then for the real-world dataset (Table \ref{tab:sisec}), after which a brief discussion of the results is given.

\subsection{\label{subsec:Results} Results on the Simulated Dataset}
Table \ref{tab:rev} shows the RMSE and MD results obtained for the the different DoA approaches at four different reverberation times based on the simulated dataset.  From the table, it can be observed that the performance of all the approaches is similar at low reverberation times of 0.0 s and 0.1 s. At the two larger reverberation times, the performances of SFF, HE-LP and GCC-PHAT are comparable, and they are better than SRP-PHAT and GCC. 

\begin{table*}[]
\centering
\caption{RMSE and  MD (in percentage) for the five sound source DoA estimation techniques by varying the reverberation time using clean speech signals from the simulated dataset. Techniques include both narrowband (NB) and broadband (BB) approaches. 
}
\label{tab:rev}
\begin{tabular}{llclclclclc}
\hline
\multicolumn{1}{|c|}{\multirow{3}{*}{\textbf{Rev. time}}} & \multicolumn{2}{c|}{\multirow{2}{*}{\textbf{\begin{tabular}[c]{@{}c@{}}SRP-PHAT\\ NB\end{tabular}}}} & \multicolumn{2}{c|}{\multirow{2}{*}{\textbf{\begin{tabular}[c]{@{}c@{}}GCC\\  BB\end{tabular}}}} & \multicolumn{2}{c|}{\multirow{2}{*}{\textbf{\begin{tabular}[c]{@{}c@{}}HE-LP\\  BB\end{tabular}}}} & \multicolumn{2}{c|}{\multirow{2}{*}{\textbf{\begin{tabular}[c]{@{}c@{}}GCC-PHAT\\  BB\end{tabular}}}} & \multicolumn{2}{c|}{\multirow{2}{*}{\textbf{\begin{tabular}[c]{@{}c@{}}SFF\\  NB\end{tabular}}}} \\
\multicolumn{1}{|c|}{} & \multicolumn{2}{c|}{} & \multicolumn{2}{c|}{} & \multicolumn{2}{c|}{} & \multicolumn{2}{c|}{} & \multicolumn{2}{c|}{} \\ \cline{2-11} 
\multicolumn{1}{|c|}{} & \multicolumn{1}{l|}{\textbf{RMSE}} & \multicolumn{1}{l|}{\textbf{MD}} & \multicolumn{1}{l|}{\textbf{RMSE}} & \multicolumn{1}{l|}{\textbf{MD}} & \multicolumn{1}{l|}{\textbf{RMSE}} & \multicolumn{1}{l|}{\textbf{MD}} & \multicolumn{1}{l|}{\textbf{RMSE}} & \multicolumn{1}{l|}{\textbf{MD}} & \multicolumn{1}{l|}{\textbf{RMSE}} & \multicolumn{1}{l|}{\textbf{MD}} \\ \hline
\multicolumn{11}{l}{} \\ \hline
\multicolumn{1}{|l|}{\textbf{0.0 s}} & \multicolumn{1}{l|}{0.94} & \multicolumn{1}{c|}{0} & \multicolumn{1}{l|}{1.70} & \multicolumn{1}{c|}{0} & \multicolumn{1}{l|}{1.70} & \multicolumn{1}{c|}{0} & \multicolumn{1}{l|}{1.70} & \multicolumn{1}{c|}{0} & \multicolumn{1}{l|}{1.70} & \multicolumn{1}{c|}{0} \\ \hline
\multicolumn{1}{|l|}{\textbf{0.1 s}} & \multicolumn{1}{l|}{1.34} & \multicolumn{1}{c|}{0} & \multicolumn{1}{l|}{1.94} & \multicolumn{1}{c|}{0} & \multicolumn{1}{l|}{1.70} & \multicolumn{1}{c|}{0} & \multicolumn{1}{l|}{1.39} & \multicolumn{1}{c|}{0} & \multicolumn{1}{l|}{1.70} & \multicolumn{1}{c|}{0} \\ \hline
\multicolumn{1}{|l|}{\textbf{0.2 s}} & \multicolumn{1}{l|}{1.34} & \multicolumn{1}{c|}{5} & \multicolumn{1}{l|}{1.98} & \multicolumn{1}{c|}{16} & \multicolumn{1}{l|}{1.70} & \multicolumn{1}{c|}{0} & \multicolumn{1}{l|}{1.34} & \multicolumn{1}{c|}{0} & \multicolumn{1}{l|}{1.71} & \multicolumn{1}{c|}{1} \\ \hline
\multicolumn{1}{|l|}{\textbf{0.3 s}} & \multicolumn{1}{l|}{1.39} & \multicolumn{1}{c|}{27} & \multicolumn{1}{l|}{2.05} & \multicolumn{1}{c|}{35} & \multicolumn{1}{l|}{1.67} & \multicolumn{1}{c|}{0} & \multicolumn{1}{l|}{1.34} & \multicolumn{1}{c|}{0} & \multicolumn{1}{l|}{1.77} & \multicolumn{1}{c|}{9} \\ \hline
\end{tabular}
\end{table*}

Table \ref{tab:noise} shows the obtained RMSE and MD values for the different methods when the speech data was corrupted with white additive noise of different SNRs. It can be observed that the MD values are high for HE-LP and SRP-PHAT. It can also be seen that although GCC-PHAT and GCC show the best performance, their results are comparable with those obtained by the SFF-based method. 

\begin{table*}[]
\centering
\caption{RMSE and  MD (in percentage) for the five sound source DoA estimation techniques using speech signals from the simulated dataset that were corrupted by additive white noise in different SNR categories. Techniques include both narrowband (NB) and broadband (BB) approaches. 
}
\label{tab:noise}
\begin{tabular}{llclclclclc}
\hline
\multicolumn{1}{|c|}{\multirow{3}{*}{\textbf{\begin{tabular}[c]{@{}c@{}}SNR\\ \end{tabular}}}} & \multicolumn{2}{c|}{\multirow{2}{*}{\textbf{\begin{tabular}[c]{@{}c@{}}SRP-PHAT\\ NB\end{tabular}}}} & \multicolumn{2}{c|}{\multirow{2}{*}{\textbf{\begin{tabular}[c]{@{}c@{}}GCC\\  BB\end{tabular}}}} & \multicolumn{2}{c|}{\multirow{2}{*}{\textbf{\begin{tabular}[c]{@{}c@{}}HE-LP\\  BB\end{tabular}}}} & \multicolumn{2}{c|}{\multirow{2}{*}{\textbf{\begin{tabular}[c]{@{}c@{}}GCC-PHAT\\  BB\end{tabular}}}} & \multicolumn{2}{c|}{\multirow{2}{*}{\textbf{\begin{tabular}[c]{@{}c@{}}SFF\\  NB\end{tabular}}}} \\
\multicolumn{1}{|c|}{} & \multicolumn{2}{c|}{} & \multicolumn{2}{c|}{} & \multicolumn{2}{c|}{} & \multicolumn{2}{c|}{} & \multicolumn{2}{c|}{} \\ \cline{2-11} 
\multicolumn{1}{|c|}{} & \multicolumn{1}{l|}{\textbf{RMSE}} & \multicolumn{1}{l|}{\textbf{MD}} & \multicolumn{1}{l|}{\textbf{RMSE}} & \multicolumn{1}{l|}{\textbf{MD}} & \multicolumn{1}{l|}{\textbf{RMSE}} & \multicolumn{1}{l|}{\textbf{MD}} & \multicolumn{1}{l|}{\textbf{RMSE}} & \multicolumn{1}{l|}{\textbf{MD}} & \multicolumn{1}{l|}{\textbf{RMSE}} & \multicolumn{1}{l|}{\textbf{MD}} \\ \hline
\multicolumn{11}{l}{} \\ \hline
\multicolumn{1}{|l|}{\textbf{0 dB}} & \multicolumn{1}{l|}{0.96} & \multicolumn{1}{c|}{5} & \multicolumn{1}{l|}{1.70} & \multicolumn{1}{c|}{0} & \multicolumn{1}{l|}{1.58} & \multicolumn{1}{c|}{5} & \multicolumn{1}{l|}{1.38} & \multicolumn{1}{c|}{0} & \multicolumn{1}{l|}{1.65} & \multicolumn{1}{c|}{4} \\ \hline
\multicolumn{1}{|l|}{\textbf{-5 dB}} & \multicolumn{1}{l|}{1.06} & \multicolumn{1}{c|}{29} & \multicolumn{1}{l|}{1.66} & \multicolumn{1}{c|}{5} & \multicolumn{1}{l|}{1.97} & \multicolumn{1}{c|}{15} & \multicolumn{1}{l|}{1.86} & \multicolumn{1}{c|}{4} & \multicolumn{1}{l|}{1.54} & \multicolumn{1}{c|}{10} \\ \hline
\multicolumn{1}{|l|}{\textbf{-8 dB}} & \multicolumn{1}{l|}{0.98} & \multicolumn{1}{c|}{57} & \multicolumn{1}{l|}{1.61} & \multicolumn{1}{c|}{10} & \multicolumn{1}{l|}{2.57} & \multicolumn{1}{c|}{26} & \multicolumn{1}{l|}{2.18} & \multicolumn{1}{c|}{11} & \multicolumn{1}{l|}{1.81} & \multicolumn{1}{c|}{15} \\ \hline
\multicolumn{1}{|l|}{\textbf{-10 dB}} & \multicolumn{1}{l|}{1.33} & \multicolumn{1}{c|}{70} & \multicolumn{1}{l|}{1.73} & \multicolumn{1}{c|}{13} & \multicolumn{1}{l|}{2.91} & \multicolumn{1}{c|}{37} & \multicolumn{1}{l|}{2.20} & \multicolumn{1}{c|}{11} & \multicolumn{1}{l|}{2.03} & \multicolumn{1}{c|}{19} \\ \hline
\end{tabular}
\end{table*}

\subsection{\label{subsec:Results1} Results on the Real-world Dataset}

The metrics obtained using the real-world SiSEC data in the comparison of the five DoA methods are shown in Table \ref{tab:sisec}.  From the table, it can be observed that SFF, HE and GCC-PHAT perform better than GCC and SRP-PHAT.

\begin{table*}
\centering
\caption{RMSE and  MD (in percentage) for the five sound source DoA estimation techniques using speech signals from the real-world SiSEC dataset. Techniques include both narrowband (NB) and broadband (BB) approaches.}
\label{tab:sisec}
\begin{tabular}{llllllllll}
\hline
\multicolumn{2}{|c|}{\multirow{2}{*}{\textbf{\begin{tabular}[c]{@{}c@{}}SRP-PHAT\\  NB\end{tabular}}}} & \multicolumn{2}{c|}{\multirow{2}{*}{\textbf{\begin{tabular}[c]{@{}c@{}}GCC\\  BB\end{tabular}}}} & \multicolumn{2}{c|}{\multirow{2}{*}{\textbf{\begin{tabular}[c]{@{}c@{}}HE-LP\\  BB\end{tabular}}}} & \multicolumn{2}{c|}{\multirow{2}{*}{\textbf{\begin{tabular}[c]{@{}c@{}}GCC-PHAT\\  BB\end{tabular}}}} & \multicolumn{2}{c|}{\multirow{2}{*}{\textbf{\begin{tabular}[c]{@{}c@{}}SFF\\  NB\end{tabular}}}} \\
\multicolumn{2}{|c|}{} & \multicolumn{2}{c|}{} & \multicolumn{2}{c|}{} & \multicolumn{2}{c|}{} & \multicolumn{2}{c|}{} \\ \hline
\multicolumn{1}{|l|}{\textbf{RMSE}} & \multicolumn{1}{l|}{\textbf{MD}} & \multicolumn{1}{l|}{\textbf{RMSE}} & \multicolumn{1}{l|}{\textbf{MD}} & \multicolumn{1}{l|}{\textbf{RMSE}} & \multicolumn{1}{l|}{\textbf{MD}} & \multicolumn{1}{l|}{\textbf{RMSE}} & \multicolumn{1}{l|}{\textbf{MD}} & \multicolumn{1}{l|}{\textbf{RMSE}} & \multicolumn{1}{l|}{\textbf{MD}} \\ \hline
 &  &  &  &  &  &  &  &  &  \\ \hline
\multicolumn{1}{|l|}{1.5} & \multicolumn{1}{l|}{20} & \multicolumn{1}{l|}{1.89} & \multicolumn{1}{c|}{17} & \multicolumn{1}{l|}{1.70} & \multicolumn{1}{c|}{0} & \multicolumn{1}{l|}{1.65} & \multicolumn{1}{c|}{0} & \multicolumn{1}{l|}{1.71} & \multicolumn{1}{c|}{3} \\ \hline
\end{tabular}
\end{table*}

 By combining the main results shown in Tables \ref{tab:rev}, \ref{tab:noise} and \ref{tab:sisec}, it can be concluded that the proposed SFF-based DoA method shows a clearly better performance than the narrowband SRP-PHAT in all the cases. Comparing the SFF-based DoA estimator with the BB estimators, we observe that while GCC performs poorly in the reverberation dataset, HE-LP performs poorly in the noisy dataset. The performance of the SFF-based DoA estimator is consistent in both of the datasets. Furthermore, the performance of the SFF-based method is only slightly lower than the performance of the state-of-the-art  GCC-PHAT method.

\subsection{Discussion} The results are in line with the known pros and cons of the approaches compared. GCC is known to be vulnerable to multiple reflections and therefore it does not perform well in conditions where speech is distorted by reverberations \cite{2012_comparison_gcc}. On the other hand, GCC with the PHAT weighting is known to give good DoA estimates in conditions with reverberation and low SNR  \cite{2008WhyPhatWorksInNoise}. SRP-PHAT is most suitable for closely spaced microphones where the effect of spatial aliasing is smaller and hence it is understandable that the method performs poorly in other, more difficult cases. The performance of HE mainly depends on the performance of LP analysis. LP analysis is known to be vulnerable to low SNR conditions \cite{1999_enhance_LP_yegna}. The proposed SFF approach benefits from robust estimation of speech regions in the SFF spectrum and by restricting the DoA estimation to these regions. However, the spectral flatness measure is not suitable for speech detection in coloured noise. To improve the SFF-based approach, other measures that can be used for the estimation of speech segments in coloured noise can be explored \cite{drygajlo_2001_spectral_flatness}.    

\section{ Summary and Conclusions}
\label{sec:conclusion}
For the DoA estimation of sound sources, a narrowband method of exploiting signal regions of high SNR in voiced speech segments based on SFF spectral envelopes was presented. The performance of the proposed SFF-based NB technique was compared to four state-of-the-art DoA estimation approaches using both simulated and real-word recordings. In all the estimators only the high SNR regions are used. The results show that the SFF-based NB method was better than the SRP-PHAT NB method in both datasets, indicating the benefit of performing the VAD and DoA estimation using the SFF spectrum instead of the STFT spectrum. Furthermore, the performance of the SFF-based approach was more consistent than HE-LP and GCC in varying noise and reverberation, and was only slightly worse than that of GCC-PHAT. 

\bibliography{DoA_References,mybib,refs.bib}
\end{document}